\documentclass[aps,prb,twocolumn,longbibliography,superscriptaddress]{revtex4-1}

\usepackage{amsmath}
\usepackage{amssymb}
\usepackage{natbib}
\usepackage{graphicx}
\usepackage{bm}
\newcommand\bi{\bibitem}

\usepackage{subfigure}
\usepackage{float}
\usepackage{epsfig}
\usepackage{hyperref}
\hypersetup{
colorlinks=true,final=true,
        linkcolor=blue,
        citecolor=blue,
        filecolor=blue,
        urlcolor=blue,
}
\def\be{\begin{equation}}
\def\ee{\end{equation}}
\def\ba{\begin{eqnarray}}
\def\ea{\end{eqnarray}}

\begin{document}

\title{Chiral anomaly induced nonlinear Hall effect in multi-Weyl semimetals}
%\date{\today}
\author{Snehasish Nandy}
\thanks{Both authors SN and CZ contributed equally.}
\affiliation{Department of Physics, University of Virginia, Charlottesville, VA 22904, USA}
\author{Chuanchang Zeng}
\thanks{Both authors SN and CZ contributed equally.}
\affiliation{ Centre for Quantum Physics, Key Laboratory of Advanced Optoelectronic Quantum Architecture and Measurement(MOE), School of Physics, Beijing Institute of Technology, Beijing, 100081, China}
\author{Sumanta Tewari}
\affiliation{Department of Physics and Astronomy, Clemson University, Clemson, SC 29634, USA}

\begin{abstract}
After the experimental realization, the Berry curvature dipole (BCD) induced nonlinear Hall effect (NLHE) has attracted tremendous interest to the condensed matter community. Here, we investigate another family of Hall effect, namely, chiral anomaly induced nonlinear Hall effect (CNHE) in multi-Weyl semimetal (mWSM). In contrast to the BCD induced NLHE, CNHE appears because of the combination of both chiral anomaly and anomalous velocity due to non-trivial Berry curvature. Using the semiclassical Boltzmann theory within the relaxation time approximation, we show that, in contrast to the chiral anomaly induced linear Hall effect, the magnitude of CNHE decreases with the topological charge $n$. Interestingly, we find that unlike the case of n=1, the CNHE has different behaviors in different planes. Our prediction on the behavior of CNHE in mWSM can directly be checked in experiments.
\end{abstract}
%\pacs{}
\maketitle
\section{Introduction} 
In recent years, the three-dimensional Dirac and Weyl semimetals have attracted tremendous interest in topological condensed matter physics. Weyl semimetals (WSMs) can accommodate gapless chiral quasiparticles, known as Weyl fermions, near the touching of a pair of non-degenerate
bands (also called Weyl nodes)~\cite{Murakami_2007, Peskin_1995, Murakami2:2007, Yang:2011, Burkov1:2011, Burkov:2011, Volovik, Wan_2011, Xu:2011}. In a WSM, the non-trivial topological properties emerge due to Weyl nodes which can act as a source or sink of the Abelian Berry curvature. Each Weyl node is associated with a chirality quantum number, known as the topological charge whose strength is related to the Chern number and is quantized in integer values~\cite{Xiao_2010}. 
%In order to have a topological charge (designated by $n$)  associated with the Weyl node, WSM has to break either time-reversal symmetry (TRS) or the space inversion symmetry (IS)~\cite{Burkov:2011, Volovik, Wan_2011, Xu:2011}.

Recently a number of inversion broken and time-reversal (TR) symmetric materials such as (TaAs, MoTe$_{2}$, WTe$_{2}$) have been experimentally proposed as WSM~\cite{Lv_2015, Huang_2015, Hasan_2015, Wu_2016, Jiang_2017,Yan_2017}. Although these systems have Weyl nodes with topological charge ($n$) equal to $\pm 1$, it has been proposed that Weyl nodes with higher topological charge $n>1$ can also be realized in condensed matter systems~\cite{Xu:2011,bernevig12, hasan16, Nagaosa_2014}. These are called multi-Weyl semimetals (mWSMs). Unlike the single WSM whose dispersion is linear in momentum along all directions (i.e. isotropic dispersion), the mWSM ($n> 1$) shows natural anisotropy in dispersion. In particular, the double WSM ($n = 2$) and triple WSM ($n = 3$) depict linear dispersion along one symmetry direction and quadratic and cubic energy dispersion relations for the other two directions respectively. Using the density functional theory (DFT) calculations, it has been proposed that HgCr$_2$Se$_4$ and SrSi$_2$~\cite{Xu:2011,bernevig12, hasan16} can be the candidate 
materials for double WSM, whereas A(MoX)$_3$ (with $A=Rb$, $TI$; $X=Te$) can accommodate triple-Weyl points~\cite{zunger17}. It is important to note that only the Weyl nodes with topological charge $n \leq 3$ can be allowed in real materials due to restriction arising from discrete rotational symmetry on a lattice~\cite{bernevig12,Nagaosa_2014}. Moreover, the single WSM can be viewed as 3D analogue of graphene whereas the double WSM and triple WSM can be represented as 3D counterparts of bilayer and ABC-stacked
trilayer graphene, respectively~\cite{falko06,peres06,macdonald08}.

Weyl semimetals offer a plethora of fascinating transport properties due to the manifestation of quantum anomalies in the presence of external electromagnetic fields. Till now, chiral anomaly (also known as Adler-Bell-Jackiw anomaly) induced negative longitudinal magnetoresistance and planar Hall effect are the two most remarkable transport properties studied in theory and experiments~\cite{Kim:2014, Son:2013,Vladimir_2017, Burkov_jpcm, Burkov_2017, Nandy_2017, Nandy_2020}. In WSMs, the numbers of left-handed and right-handed Weyl fermions are separately conserved in the absence of any external gauge or gravitational
field coupling. On the other hand, this number conservation is violated in the presence of  non-orthogonal electric ($\mathbf{E}$) and magnetic ($\mathbf{B}$) fields (i.e., $\mathbf{E \cdot B \neq 0}$). This effect is known as chiral anomaly~\cite{Goswami:2013, Adler:1969, Bell:1969, Nielsen:1981, Nielsen:1983, Aji:2012, Zyuzin:2012, Volovik, Wan_2011, Xu:2011,Moore_2015}. The proposed LMR and PHE induced by the chiral anomaly  have already been realized in several experiments~\cite{He:2014, Chen_2015, Liang:2015,CLZhang:2016,QLi:2016,Xiong, Li_2018, Liang_2018, deng19, Chen_2018, Singha_2018, kumar18}. The corresponding current ($\mathbf{J}$), which is linear in electric field in both cases, can be expressed as $\mathbf{J} \propto  (\mathbf{E \cdot B})\mathbf{B}$.

Recently, another interesting transport property induced by chiral anomaly -- nonlinear Hall effect (NLHE) -- has been proposed in the context of single WSM~\cite{Burkov_2020}. This chiral anomaly induced nonlinear Hall effect (CNHE) is different from NLHE induced by Berry curvature dipole (BCD)~\cite{Inti_2015, Moore_2016} because the latter can survive in the absence of external magnetic field. The CNHE is second-order in electric field and appears due to the combination of both the chiral anomaly and the anomalous velocity ($\mathbf{v_a}$) due to Berry curvature. The corresponding current density can be expressed as $\mathbf{J}^{CN}=-e\sum_{\mathbf{k},\chi} \mathbf{v_a^{\chi}} \Delta n_{k}^{\chi}$ where $\chi$ is the chirality of the Weyl node, $\mathbf{v_a} \sim \mathbf{E \times \Omega_k}$~\cite{Xiao_2010} and $\Delta n_{k} \sim \chi \mathbf{E \cdot B}$ is the modification of the chiral electron density in the vicinity of each Weyl node. The NLHE caused by both the chiral anomaly as well as BCD have been studied in single WSMs~\cite{Burkov_2020,Yan_2018,Inti_2019,Zeng_2020} whereas they have not yet been explored in the context of mWSMs. Recent
experimental realizations~\cite{Kang_2019,QMa_2019} of BCD induced NLHE in single WSM add to the interest of experimental verification of these effects in various systems.

In this paper, we investigate the chiral anomaly induced nonlinear Hall effect in mWSMs using low-energy model. The main findings of this work are the following: Using the quasiclassical Boltzmann transport theory within the relaxation time approximation, we show that the CNHE in mWSMs can only survive in the presence of achiral tilt (i.e., tilt of the opposite chirality nodes are in same direction) of the Weyl nodes when the Weyl nodes are at same energy. On the other hand, this restriction no longer exists in WSMs with nonzero chiral chemical potential (i.e., the Weyl nodes are located at different energies). We further analytically show that the magnitude of CNHE depends non-trivially on topological charge $n$ (see Eqs.~7 and 8). Although the dependencies are nontrivial, the magnitude of CNHE decreases with $n$. This is in contrast with the case of chiral anomaly induced linear Hall effect where the magnitude increases with $n$. Interestingly, the chemical potential dependence ($\mu^{-2/n}$) remains unchanged in both linear and nonlinear cases. Moreover, we also find that the CNHE shows different behavior (i.e., different coefficients) in single and triple WSMs when external electromagnetic fields are rotated in different planes whereas they have the same behavior in the case of double WSM (see Eqs.~7 and 8). 

The rest of the paper is organized as follows. In Sec.~\ref{formalism_CNHE}, we formulate the general expression of CNHE using quasiclassical Boltzmann formalism. In Sec.~\ref{model}, we introduce the low-energy Hamiltonian of multi-Weyl semimetals. In Sec.~\ref{result}, we derive the analytical expressions of CNHE in mWSMs and find the dependencies of CNHE with topological charge $n$. Finally, we summarize our results and discuss possible future directions in Sec.~\ref{cons}.

\section{Formalism of Chiral anomaly induced NLHE} 
\label{formalism_CNHE}
In the presence of electric and magnetic fields, transport properties get substantially modified due to the presence of non-trivial Berry curvature 
which acts as a fictitious magnetic field in the momentum space. The steady state phenomenological Boltzmann transport equation within the relaxation time approximation takes the form~\cite{John_2001}
\begin{equation}
(\mathbf{\dot{r}}\cdot\mathbf{\nabla_{r}}+\mathbf{\dot{k}}\cdot\mathbf{\nabla_{k}})g_{\mathbf{k}}=\frac{g_{0}-g_{\mathbf{k}}}{\tau(\mathbf{k})},
\label{eq_BZf}
\end{equation}
where $g_{0}$ is the equilibrium Fermi-Dirac distribution function and $\tau(k)$ is the intranode scattering time assuming the internode scattering time ($\tau_s$) is much greater than intranode scattering time. We neglect internode scattering time because the terms related to internode scattering do not contribute to CNHE as shown in Ref.~\cite{Burkov_2020}. Here, $g_{\mathbf{k}}$ is the distribution function
in the presence of perturbative fields. In this work, we ignore the  momentum dependence of $\tau$ for simplifying the calculations and assume it to be a constant~\cite{Kim:2014, Son:2013,Burkov_2020}. In presence of Berry curvature as well as an electromagnetic field, the semiclassical equations of motion for an electron can be written as~\cite{Son_2012,Duval_2006}
\begin{eqnarray}
&&\mathbf{\dot{r}}=D(\mathbf{B,\Omega_{k}})[\mathbf{v_{k}}+\frac{e}{\hbar}(\mathbf{E}\times
\mathbf{\Omega_{k}})+\frac{e}{\hbar}(\mathbf{v_{k}}\cdot\mathbf{\Omega_{k}})\mathbf{B}], \nonumber \\
&&\hbar\mathbf{\dot{k}}=D(\mathbf{B,\Omega_{k}})[e\mathbf{E}+\frac{e}{\hbar}(\mathbf{v_{k}}
\times \mathbf{B})+\frac{e^{2}}{\hbar}(\mathbf{E}\cdot\mathbf{B})\mathbf{\Omega_{k}}], \nonumber \\
\label{eq_motion}
\end{eqnarray}
where $D(\mathbf{B,\Omega_{k}})=(1+\frac{e}{\hbar}(\mathbf{B}\cdot \mathbf{\Omega_{k}}))^{-1}$ is the phase space
factor as the Berry curvature  $\mathbf{\Omega_{k}}$ modifies the phase space volume element 
$dkdx \rightarrow D(\mathbf{B,\Omega_{k}})dkdx$~\cite{Duval_2006}.
%Hereafter, we denote $D(\mathbf{B,\Omega_{k}})$ by $D$. 
The term $\propto (\mathbf{E}\cdot\mathbf{B})$ is responsible for chiral anomaly which arises in axion-electrodynamics of WSM. Now plugging the above equation into Eq.~(\ref{eq_BZf}), the distribution function $g_{\mathbf{k}}$ upto second-order in $E$ for spatially uniform external fields can be obtained as $g_{\mathbf{k}}= g_0+g_1+g_2$ where

\begin{eqnarray}
&&g_1= \frac{\tau}{D(\mathbf{B,\Omega_{k}})}\left[e\mathbf{E \cdot v_k}+\frac{e^2}{\hbar}(\mathbf{E \cdot B})(\mathbf{\Omega_k \cdot v_k})\right]\left(\frac{\partial g_0}{\partial \epsilon_k}\right)\nonumber \\
&&g_2=\frac{\tau}{D(\mathbf{B,\Omega_{k}})}\left[e\mathbf{E \cdot v_k}+\frac{e^2}{\hbar}(\mathbf{E \cdot B})(\mathbf{\Omega_k \cdot v_k})\right]\left(\frac{\partial g_1}{\partial \epsilon_k}\right) \nonumber \\
\label{dis_fun}
\end{eqnarray}
From the general expression of the current density $J=-e\int [dk] D^{-1} \mathbf{\dot{r}} g_{\mathbf{k}}$, the nonlinear Hall current density upto the order of ($\tau E^2B$) in the presence of external fields can be obtained as~\cite{Burkov_2020}
\begin{eqnarray}
\mathbf{J}^{CN}&&=\sum_{s}\frac{e^4 \tau}{\hbar^2}\int \frac{d^3k}{(2\pi)^3} [(\mathbf{E \cdot v_{s,k}})(\mathbf{B \cdot \Omega_{s,k}})-(\mathbf{E \cdot B}) \nonumber \\
&&(\mathbf{v_{s,k} \cdot \Omega_{s,k}})](\mathbf{E \times \Omega_{s,k}})\left(\frac{\partial g_0^{s,k}}{\partial \epsilon_k^s}\right)
\label{non_den}
\end{eqnarray} 
The above equation indicates that the CNHE is a purely Fermi surface quantity and vanishes in an inversion-symmetric system. Here, we have ignored the higher order (i.e., $\tau^2$ dependent) contribution. 
%We would like to point out that the second term in Eq.~(\ref{non_den}) vanishes for $E \perp B$, while in mWSM the first term remains nonzero in this limit.

%It is clear from the Eq.~(\ref{non_den}) that the direction of the nonlinear Hall current is perpendicular to the plane of anomalous velocity generated by the Berry curvature. The second term of the above equation is coming from the chiral anomaly. Therefore, in the presence of $E$ and $B$ perpendicular to each other, the second term will vanish whereas the first term may still be finite. 

It is important to note that along with the chiral anomaly induced contribution, there may exist other contributions to the nonlinear Hall current such as BCD induced contribution and disorder-mediated contributions such as nonlinear side jump and skew-scattering contributions. However, the chiral anomaly induced NLHE is different from the BCD induced~\cite{Inti_2015} as well as disorder-mediated NLHE~\cite{Sodemann_2019}, both of which are independent of magnetic field. Therefore, CNHE can be separated from BCD- and disorder-induced NLHE by examining their dependence on the magnetic field~\cite{Inti_2015,Inti_2019,Zeng_2020,Pesin_2019, Sodemann_2019}. 
%Another nonlinear Hall effect in the presence of electromagnetic field has been derived for WSM~\cite{Moore_2016}, which is nonzero when the electric and magnetic fields are perpendicular to each other and therefore can be separated from CNHE in experiment. 
%The proposed CNHE can also be separated from the Hall effect induced by valley-selective Joule heating~\cite{Nandy_2020_PRL}. 
Moreover, the chiral anomaly induced nonlinear Hall effect can also be differentiated from linear Hall effects by measuring second harmonic Hall resistance in a.c experiments. 

%Now the chiral anomaly induced nonlinear Hall current can written in tensorial form as

%\begin{eqnarray}
%J^{CN}&&=\frac{e^4 \tau}{\hbar^2}\int \frac{d^3k}{(2\pi)^3} \left[(\mathbf{E \cdot v_k})(\mathbf{B \cdot \Omega_k})-(\mathbf{E \cdot B})(\mathbf{v_k \cdot \Omega_k})\right] \nonumber \\
%&&(\mathbf{E \times \Omega_k})\left(\frac{\partial f_0}{\partial \epsilon_k}\right)
%\label{non_den}
%\end{eqnarray} 

\section{Model Hamiltonian} 
\label{model}
The low-energy effective Hamiltonian describing a Weyl node with topological charge $n$ and chirality $s$ can be written as~\cite{Roy_2016, Mukherjee_2018}
\begin{eqnarray}
&&H_{n}^{s} \left( \mathbf{k} \right) \nonumber \\
&&= s\left[\alpha_{n} k^n_{\bot} \left[ \cos \left( n \phi_{k} \right) 
\sigma_{x} +\sin \left( n \phi_{k} \right) \sigma_{y} \right] + v (k_z-sQ) \sigma_{z}\right]\nonumber \\
&& + C_s v(k_z-sQ)-sQ_0
\label{eq_multi1}
\end{eqnarray}
where $k_{\bot}=\sqrt{k_x^2+k_y^2}$, $\phi_k={\rm arctan}(k_y/k_x)$ and $\sigma_i$'s $\left( \sigma_x,\sigma_y,\sigma_z \right)$ are the Pauli matrices representing the pseudo-spin indices. The Weyl nodes are shifted by an amount $\pm Q$ in momentum
space due to broken time-reversal symmetry whereas the broken inversion symmetry shifts the nodes in energy by $\pm Q_0$. Here, $\alpha_n=\frac{v_{\perp}}{k_0^{n-1}}$ where $v_{\perp}$ is the effective velocity of the quasiparticles in the plane perpendicular to the $z$ axis and $k_0$ represents a material-dependent parameter having the dimension of momentum. $v$ and $C_s$ denote the velocity and tilt parameter along the $z$-direction respectively. In this work, we restrict ourselves to type-I multi-Weyl node i.e., $|C_s| <1$ which indicates that the Fermi surface is point-like at the Weyl node. The energy dispersion of the multi-Weyl node is given by $\epsilon_{\mathbf{k}}^{\pm} =C_s (k_z-sQ)-sQ_0 \pm \sqrt{ \alpha^2_{n} k^{2 n}_{\bot} + v^2 k^2_z}$  where $\pm$ represents conduction and valence bands respectively. It is now clear that the dispersion around a Weyl node with $n=1$ is isotropic in all momentum directions. On the other hand, for $n>1$, we find that the dispersion 
around a double (triple) Weyl node becomes quadratic (cubic) along both $k_x$ and $k_y$ directions whereas varies linearly with $k_z$. Now, the different Berry curvature components associated with the multi-Weyl node are given by
\begin{equation}
{\Omega}_{s,\mathbf{k}}^{\pm} =\pm \frac{s}{2} \frac{n v \alpha_n^2 k^{2n-2}_{\bot} }{\epsilon_{\mathbf{k}}^{3}}
\: \{ k_x, k_y, n (k_z-sQ) \}.
\label{eq_bcl}
\end{equation}
It is clear from the Eq.~(\ref{eq_bcl}) that, similar to energy dispersion, the Berry curvature is isotropic in all momentum directions for single Weyl case whereas becomes anisotropic for WSMs with $n>1$ i.e., for double WSM ($n=2$) and triple WSM ($n=3$) due to the presence of $k^{2n-2}_{\bot}$ factor and monopole charge $n$.  
The above observation itself is an indication that the multi-Weyl nature can indeed modify CNHE, which appears due to combination of both chiral anomaly and anomalous velocity induced by non-trivial Berry curvature, in double and triple WSMs as compared to single Weyl case.

\section{Results}
\label{result}
To calculate the different components of chiral anomaly induced nonlinear current (CNC), we apply in-plane ($xy$ plane) electric and magnetic fields. To perform the integration (Eq.~\ref{non_den}) for type-I multi-Weyl semimetal, we use cylindrical coordinate geometry and make several transformations--(i) $k_{\perp}=k_{\perp}\alpha_n^{-1/n}, \quad k_z=k_z/v$; (ii) $k_{\perp}=k_{\perp}^{-1/n}$; (iii) $k_{\perp}=k \sin \theta, \quad k_z=k \cos \theta$. After some algebra, $\mathbf{J}^{CN}$ at zero temperature can be obtained as
\begin{eqnarray}
[\mathbf{J}^{CN}]_{xy}= && \frac{e^4 \tau \alpha_n^{\frac{2}{n}}}{64\pi^{\frac{3}{2}}\hbar^2} \frac{n[n (9 + 2 n)-2]\Gamma[2 - \frac{1}{n}]}{ \Gamma[\frac{9}{2} - \frac{1}{n}]}\sum_s \frac{C_s}{\mu_{s}^{2/n}}\nonumber \\
&&(\mathbf{E \cdot B}) (\mathbf{\hat{z} \times E}) \nonumber \\
%[\mathbf{J}^{CN}]_{xy}&&=\frac{e^4 \tau n^3 \alpha^{\frac{2}{n}}}{16\pi^2\hbar^2} \frac{[n (9 + 2 n)-2]\sqrt{\pi} \Gamma[2 - \frac{1}{n}]}{4 n^2 \Gamma[\frac{9}{2} - \frac{1}{n}]}\sum_s \frac{C_s}{\mu_s^{2/n}}\nonumber \\
%&&(\mathbf{E \cdot B}) (\mathbf{\hat{z} \times E}) \nonumber \\
\label{non_den_xy_mwsm}
\end{eqnarray}
where we have added the contribution of two nodes of opposite chirality. Here, $\mu_s=\mu+sQ_0$. It is clear from the Eq.~(\ref{non_den_xy_mwsm}) that the nonlinear current is restricted in the same plane ($xy$) with the applied fields and flows perpendicular to the tilt direction ($z$-direction in current study). Therefore, we can define this effect as \textit{chiral anomaly induced nonlinear planar Hall effect}. 

We now first consider $Q_0=0$ i.e., the Weyl nodes are at same energies. In this case, it is evident from the above equation that the chiral anomaly induced nonlinear planar Hall current (CNPHC) becomes only finite when the sign of the tilt of opposite chirality nodes are same ($C_+=C_-$). In other words, the CNPHC vanishes in the absence of tilt ($C_+=C_-=0$) and even in the presence of chiral tilt ($C_+=-C_-$) of the Weyl node. 
%Although the CNPHC vanishes in type-I chiral tilted mWSMs, by subtracting contributions coming from opposite chirality nodes, one can indeed observe a tilt-induced nonlinear Hall current $\mathbf{J}^{TN}$ (i.e., $\mathbf{J}^{TN}=\mathbf{J}^{CN}_{+}-\mathbf{J}^{CN}_{-}$). 
Considering $Q_0 \neq 0$ i.e., when the Weyl nodes are located in different energy, one can see from the Eq.~(\ref{non_den_xy_mwsm}) that the condition for finite CNPHC changes. In this case, the CNPHC can be non-zero in both cases i.e., for chiral as well as achiral tilt configurations of the Weyl nodes.

From the Eq.~(\ref{non_den_xy_mwsm}), it is clear that the magnitude of chiral anomaly induced nonlinear Hall current depends non-trivially on topological charge $n$. Although the dependencies is nontrivial, the magnitude of CNPHC decreases with $n$. This is in contrast with the case of linear planar Hall effect where the magnitude increases with $n$\cite{Nag_2020}. Moreover, we also find that the multi-Weyl nature (i.e., the $n$ dependence) also comes into CNPHC through chemical potential as $\mu^{-2/n}$. Interestingly, the chemical potential dependence ($\mu^{-2/n}$) remains unchanged in both linear and nonlinear cases \cite{Nag_2020}. These scaling factors of CNPHC with the topological charge might distinguish a single, double and triple WSM from each other in experiment. Having done a more detailed analysis, surprisingly we find that in the presence of non-orthogonal external fields (i.e., $\mathbf{E \cdot B} \neq 0$), although the second term of Eq.~(\ref{non_den}) is contributing to CNPHC for all WSMs, the first term vanishes in the case of double WSM.
%Although the chiral tilt of the Weyl nodes force to vanish total CNPHC (sum of the contribution of two nodes), it can generate chirality-dependent nonlinear planar Hall current (difference of the contribution coming from two nodes) which can be detected via circular dichroism.

We would like to point out the most striking difference between the BCD induced and chiral anomaly induced nonlinear Hall effect. In the case of BCD induced nonlinear Hall effect, the contribution associated with the Weyl nodes vanishes (irrespective of the presence or absence of the tilt of the Weyl nodes). On the other hand, the whole contribution of the chiral anomaly induced nonlinear Hall effect is generated from the Weyl nodes. 

Now we study the CNHE when the external fields are restricted in $xz$ plane. In this configuration, $\mathbf{J}^{CN}$ at zero temperature can be written as
\begin{eqnarray}
&&[\mathbf{J}^{CN}]_{xz}=\frac{e^4 \tau n \alpha_n^{\frac{2}{n}}}{64\pi^{\frac{3}{2}}\hbar^2} [\frac{[n (9 + 2 n)-2]\Gamma[2 - \frac{1}{n}]E_x B_x}{ \Gamma[\frac{9}{2} - \frac{1}{n}]}\nonumber \\
&&+\frac{48n^5\Gamma[3 - \frac{1}{n}] E_z B_z}{(7n-2)(5n-2)(3n-2)\Gamma[\frac{1}{2} - \frac{1}{n}]}]\sum_s \frac{C_s}{\mu_s^{2/n}}(\mathbf{\hat{z} \times E_x}) \nonumber \\
\label{non_den_xz_mwsm}
\end{eqnarray}
The above equation suggests that unlike $xy$ plane, the nonlinear Hall current flows perpendicular to the plane containing external electric and magnetic fields (i.e., along $y$ direction). Comparing Eq.~(7) and Eq.~(8), it is clear that in the case of $[\mathbf{J}^{CN}]_{xz}$, the coefficients proportional to $E^2_x B_x$ and $E_x E_z B_z$ are different whereas the coefficients proportional to $E^2_x B_x$ and $E_x E_y B_y $ for $J^{CN}_{xy}$ are same. This leads to the fact that the current in the $y$ direction in the presence of external electromagnetic fields restricted in $xy$ plane will be different when the electromagnetic fields are rotated in $xz$ plane. This fact gives rise to the planar anisotropy for chiral anomaly induced nonlinear Hall effect. Interestingly, we find that this planar anisotropy is present in the case of single and triple WSMs whereas this is no longer exists in double WSM.

From the Eq.~(\ref{non_den_xz_mwsm}), it is clear that similar to the $xy$ plane, the chiral anomaly induced nonlinear Hall current depends non-trivially on topological charge $n$. In particular, the magnitude of CNPHC decreases as we go from single WSM to higher order WSMs ($n>1$) at a fixed chemical potential. Although the topological charge dependence of CNPHC is different for $xz$ plane compared to $xy$ plane, the chemical potential dependence ($\mu^{-2/n}$) remains unchanged in both planes. 

The analytical results obtained in Eqs.~(7), (8) agree very well with the numerical calculations, as shown in Fig.~1. Note that, $J^{CN}_{xy}$ has same coefficients proportional to $E^2_x B_x$ and $E_x E_y B_y $ while $J^{CN}_{xz}$ shows different coefficients proportional to $E^2_xB_x$ and $E_x E_z B_z$. To illustrate this point more clearly, we plot the magnitude of $J^{CN}_{xz}$ as a function of angle $\theta$, the angle formed between the projections of the electric and magnetic fields on the $xz$-plane (i.e., $\left<\bm{E}, \bm{B}\right>=\theta$). As shown in Fig.~2, $J^{CN}_{xz}(\theta=\pi/2, 3\pi/2)$ equals to zero for the case of $n=2$, while $J^{CN}_{xz} (n=3)$ shows an obvious anisotropy among its $x$-direction ($\theta=0, \pi/2$, the first term in Eq.~(8)) and $z$-direction ($\theta=\pi/2, 3\pi/2$,the second term in Eq.~(8)) contribution, whose magnitudes are implied by the blue and red solid lines respectively. Interestingly, a close look into Eqs.~(7) and (8) suggests that the magnitude as well as $n$ dependence of nonlinear Hall current along the $y$ direction is different for single and triple WSMs whereas remains same for double WSM. 
%%%%%%%%%%%%%%%%%%%%%%%%%%%%%%%%%%%%%%%%%%%%%%%%%%% FIG 1
\begin{figure}[!tp]
	\begin{center}
		\includegraphics[width=0.46\textwidth]{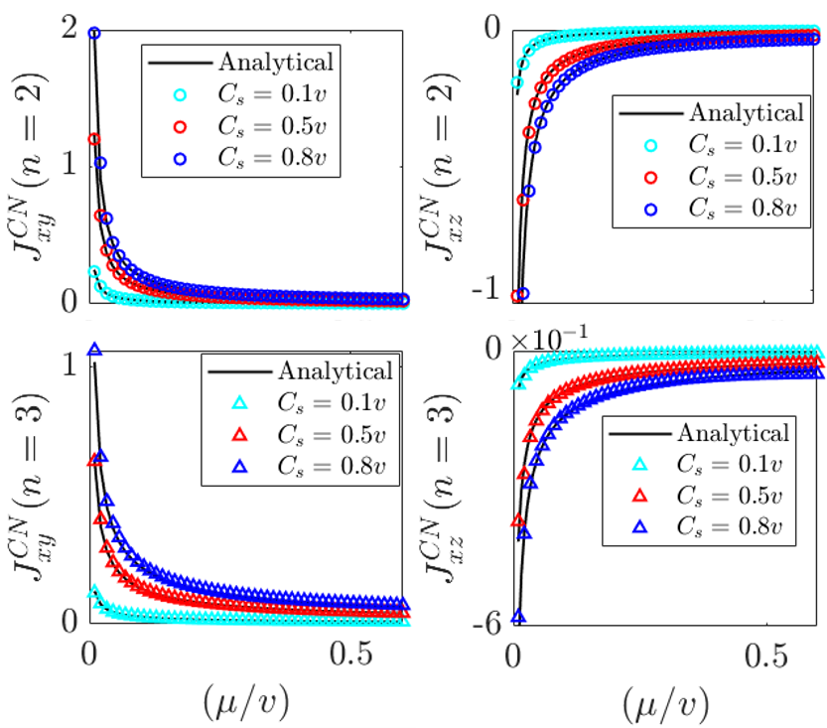}
		\llap{\parbox[b]{163mm}{\large\textbf{(a)}\\\rule{0ex}{67mm}}}
     	\llap{\parbox[b]{164mm}{\large\textbf{(b)}\\\rule{0ex}{35mm}}}
     	\llap{\parbox[b]{82mm}{\large\textbf{(c)}\\\rule{0ex}{67mm}}}
     	\llap{\parbox[b]{84mm}{\large\textbf{(d)}\\\rule{0ex}{35mm}}}
	\end{center} 
	%\vspace{-5mm}
	\caption{(Color online) Chiral anomaly induced nonlinear Hall current $\bm{J}^{CN}$ as a function of chemical potential with different tilt strength ($C_s =0.1, 0.5, 0.8v$) for a single multi-Weyl node. Panel (a),(b) show the component $\bm{J}^{CN}_{xy}$ of the nonlinear Hall current with $n=2$ and $n=3$ respectively, while panel (c), (d) show the component $\bm{J}^{CN}_{xz}$ of the nonlinear Hall current with $n=2$ and $n=3$ respectively. The symbols represent numerical results calculated directly from Eq.~(4), while the corresponding black lines indicate the analytical results based on Eq.~(7),(8). Here we use $v=0.37~eV \AA, v_{\perp}=0.32~eV \AA, k_0=0.8\AA, Q=Q_0=0$. }
	\label{fig:f1}
\end{figure}
%%%%%%%%%%%%%%%%%%%%%%%%%%%%%%%%%%%%%%

%%%%%%%%%%%%%%%%%%%%%%%%%%%%%%%%%%%%%%%%%%%%%%%%%%% FIG 2
\begin{figure}[!tp]
	\begin{center}
		\includegraphics[width=0.46\textwidth]{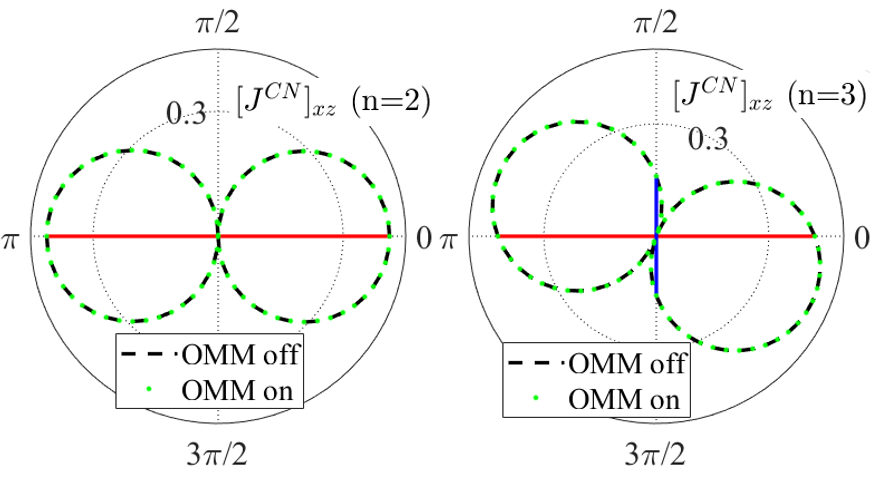}
		\llap{\parbox[b]{163mm}{\large\textbf{(a)}\\\rule{0ex}{38mm}}}
     	\llap{\parbox[b]{78mm}{\large\textbf{(b)}\\\rule{0ex}{38mm}}}
	\end{center} 
	%\vspace{-5mm}
	\caption{(Color online) The polar plot of chiral anomaly induced nonlinear Hall current $\bm{J}^{CN}_{xz}$ as a function of field angle $\theta$ (where $\theta$ is the angle between the components of electric and magnetic fields on the $xz$-plane). Panel (a) and (b) show the polar distribution $\bm{J}^{CN}_{xz}(\theta)$ for $n=2$ and $n=3$ respectively. The solid red and blue lines indicate the magnitudes of $\bm{J}^{CN}_{xz}$ proportional to $E_xB_x$ ($\theta=0, \pi$) and $E_z B_z$ ($\theta =\pi/2, 3\pi/2$), respectively. The different lengths of the blue and red lines indicate the anisotropy of the chiral anomaly induced nonlinear Hall effect in multi-Weyl systems. Note that, the factor $(\bm{\hat{z}\times E}_{x})$ as given in Eq~(8) is perpendicular to the $xz$-plane and is ignored here. We also consider the effect of orbital magnetic moment (OMM) on the magnitude of $J^{CN}_{xz}$,the numerical results are represented by the green dots.}
	\label{fig:f2}
\end{figure}
%%%%%%%%%%%%%%%%%%%%%%%%%%%%%%%%%%%%%%

%\begin{eqnarray}
%&&[\mathbf{J}^{CN}_{check}]_{xz}=\frac{e^4 \tau n \alpha^{\frac{2}{n}}}{64\pi^{\frac{3}{2}}\hbar^2}\sum_s \frac{C_s}{\mu_s^{2/n}} (\mathbf{\hat{z} \times E_x})\nonumber \\
%&&\left[\frac{\Gamma[2 - \frac{1}{n}]E_x B_x}{ \Gamma[\frac{9}{2} - \frac{1}{n}]}+\frac{2n(n-2)\Gamma[3 - \frac{1}{n}] E_z B_z}{\Gamma[\frac{9}{2} - \frac{1}{n}]}\right] \nonumber \\
%\label{non_den_xz_mwsm_test}
%\end{eqnarray}
%In addition, the quasi-particle velocity ($v_{\mathbf{k}}=\frac{\partial \epsilon_{\mathbf{k}}}{\partial \mathbf{k}}$) associated with the multi-Weyl node is given by
%\be
%v_{\mathbf{k}}=\frac{1}{\epsilon_{\mathbf{k}}}(k_x n \alpha_n^2 k_{\bot}^{2(n-1)},k_y n \alpha_n^2 k_{\bot}^{2(n-1)},v^2 k_z).
%\label{eq_vel}
%\ee
%It is clear from Eq.~(\ref{eq_vel}) that the velocity for a single WSM shows the isotropic nature in all
%momentum directions. One can figure out from the same equation that the velocity is no longer isotropic if
%we consider WSMs with $n>1$ compared to the single WSM. In particular, since the energy dispersion becomes anisotropic 
%in double and triple WSMs as described in Eq.~(\ref{eq_multi2}), the $x$ and $y$ components of the velocity vary
%with different power in $k_x$ and $k_y$ due to the factor $k_{\bot}^{2(n-1)}$ in these cases  while 
%$v_z$ remains unaltered (varies linearly with $k_z$) irrespective of the value of $n$.

The wave packet of a Bloch electron carries an orbital magnetic moment (OMM) in addition to its spin moment due to the self rotation around its center of mass. The orbital moment $m(\mathbf{k})$ which couples to the magnetic field (B) through a Zeeman-like term $\mathbf{m(k) \cdot B}$, modifies the unperturbed band energy and quasiparticle group velocity as $\tilde{\epsilon_k}=\epsilon_k-\mathbf{m \cdot B}$ and $\tilde{v_\mathbf{k}}=v_\mathbf{k}-\frac{1}{\hbar}\nabla(\mathbf{m \cdot B})$ and consequently, semiclassical equations of motion. The orbital magnetic moment for a multi-Weyl node is given by

\begin{equation}
{m}_{s,\mathbf{k}}^{\pm} = \frac{se n v \alpha_n^2 k^{2n-2}_{\bot} }{2\hbar\epsilon_{\mathbf{k}}^{2}}
\: \{ k_x, k_y, n (k_z-sQ) \}.
\label{eq_om}
\end{equation}
It is clear from the Eq.~(\ref{eq_om}) that the orbital magnetic moment is anisotropic in mWSMs compared to single WSM. In order to calculate chiral anomaly induced nonlinear Hall current numerically in mWSMs, we have chosen $k_0=1,v=0.5,v_{\perp}=1, |C_s|=0.5, Q_0=0.5, Q=2.5$, $k_0=0.8,v=0.37~eV \AA,v_{\perp}=0.32~eV\AA, |C_s|=0.8, Q_0=0, Q=0$. We find that the presence of orbital magnetic moment does not affect the magnitude of $J^{CN}$, as shown in Fig.~2. This feature, distinct from that in the anomalous responses in the linear regime \cite{Nandy_2018, Das_2019}, in turn can also be used to distinguish the chiral anomaly induced nonlinear Hall effect from linear Hall effects in experiments.

%===================================================================================
%\begin{figure}[htb]
%\centering
%\includegraphics[width=8.5cm]{OMM.pdf} 
%\caption{(Color online) We have taken  }
%\label{CNPHE_OMM}
%\end{figure}
%==================================================================
\section{Discussion and Conclusion}
\label{cons}
We investigate the chiral anomaly induced nonlinear Hall effect in multi-Weyl semimetals. In the presence of non-orthogonal electromagnetic fields, it appears because of the combination of both chiral anomaly and anomalous velocity due to non-trivial Berry curvature in WSM. Using the quasiclassical Boltzmann theory within the relaxation time approximation, we have predicted the behavior of CNHE considering low-energy model of type-I mWSMs, specifically, using two separate multi-Weyl nodes of opposite chiralities in the presence of external electric and magnetic fields rotating in the i) $xy$ plane and ii) $xz$ plane. 

We find that the chiral anomaly induced nonlinear Hall current flows perpendicular to the tilt direction i.e., perpendicular to the $z$ direction in the present work. In both cases, we show that, when the Weyl nodes are located at same energy, the CNHE in mWSMs can only be nonzero in the presence of achiral tilt (i.e., tilt of the opposite chirality nodes are in same direction) of the Weyl nodes. Interestingly, this restriction no longer exists when the Weyl nodes are located at different energies in WSM (i.e., in the presence of a nonzero chiral chemical potential). We further analytically show that, in both cases, the magnitude of CNHE depends non-trivially on topological charge $n$ (see Eqs.~7 and 8). Although the dependencies is nontrivial, the magnitude of CNHE decreases with $n$. This is in contrast with the case of chiral anomaly induced linear Hall effect where the magnitude increases with $n$. Interestingly, the chemical potential dependence ($\mu^{-2/n}$) remains unchanged in both linear and nonlinear cases. 

We find that the CNHE shows different behavior (i.e., different coefficients) when external electromagnetic fields are rotated in different planes (see Eqs.~7 and 8). Specifically, we find that the current in the $y$ direction with the external electromagnetic fields rotated in $xy$ plane will be different than the case when the external electromagnetic fields are rotated in $xz$ plane. This fact gives rise to the planar anisotropy for chiral anomaly induced nonlinear Hall effect. Interestingly, we find that this planar anisotropy is present in the case of single and double WSMs whereas this is no longer exists in double WSM. Therefore, the CNHE can be used as a probe to distinguish single, double and triple WSMs from each other in experiments. We also find that unlike the linear response case, the orbital magnetic moment does not affect on CNHE.

In contrast to the linearized model we used in this work, a real mWSM may contain Weyl nodes with different tilt with respect to one another as well as number of pair of nodes can be greater than one. In addition, the CNHE calculated using the low energy model becomes dependent on the momentum cutoff in the case of type-II mWSMs~\cite{Burkov_2020}. On the other hand, we know that a lattice model of Weyl fermions with lattice regularization provides a natural ultra-violet cut-off to the low-energy Dirac spectrum. Therefore, to predict the correct experimental behavior of CNHE in mWSMs, one needs to study a mWSM Hamiltonian using DFT or a  lattice Hamiltonian of an inversion broken mWSM. Finding a lattice description of an inversion broken mWSM and calculation of CNHE are interesting questions which we leave for future study. Investigating CNHE in the quantum regime (high magnetic field), where the Landau level quantization is applicable, would also be a fascinating question to look into. Similar to chiral anomaly induced linear Nernst effect~\cite{Sharma:2016, Nag_2020}, we also expect a finite nonlinear Nernst effect induced by chiral anomaly in mWSMs, which is yet to be explored.

\begin{acknowledgements}
S. N. acknowledges the National Science Foundation Grant No. DMR-1853048. C. Z. acknowledges support from the National Key R$\&$D Program of China (Grant No. 2020YFA0308800). S. T. thanks the
NSF-QIS-18-564 for support.
\end{acknowledgements}

%\bibliography{NLHE_mWSM}
%\bibliographystyle{apsrev}

\end{document}